\documentclass[prl,
showpacs,
superscriptaddress,
preprintnumbers,
amsmath,
amssymb,
aps,onecolumn]{revtex4-2} 

\usepackage{amsmath}  
\usepackage{amsfonts} 
\usepackage{graphicx} 
\usepackage{subfigure}
\usepackage{epstopdf}
\usepackage{color}
\usepackage{threeparttable}
\usepackage{float}
\usepackage[sort&compress]{natbib}
\usepackage{xcite} 
\usepackage{setspace}

\usepackage{titlesec}
\titleformat{\section}[block]{\normalfont\large\bfseries}{\thesection}{1em}{} 
\titleformat{\subsection}[block]{\normalfont\large\bfseries}{\thesubsection}{1em}{} 

\usepackage{chngcntr}  
\renewcommand{\thesection}{\arabic{section}.}
\renewcommand{\thesubsection}{\thesection\arabic{subsection}.} 

\begin{document}


\title{Data-driven characterization of spatiotemporal chaos using ensemble reservoir computing}

\author{Xiaoqi Lei}
\affiliation{School of Physical Science and Technology, Beijing University of Posts and Telecommunications, Beijing 100876, China} 

\author{Zixiang Yan}
\affiliation{School of Physical Science and Technology, Beijing University of Posts and Telecommunications, Beijing 100876, China}

\author{Jian Gao}
\email{gao.jian@bupt.edu.cn}
\affiliation{School of Physical Science and Technology, Beijing University of Posts and Telecommunications, Beijing 100876, China} 
\affiliation{State Key Laboratory of Information Photonics and Optical Communications, Beijing University of Posts and Telecommunications, Beijing 100876, China}

\author{Yueheng Lan}
\affiliation{School of Physical Science and Technology, Beijing University of Posts and Telecommunications, Beijing 100876, China} 

\author{Jinghua Xiao}
\affiliation{School of Physical Science and Technology, Beijing University of Posts and Telecommunications, Beijing 100876, China}

\date{\today}

\begin{abstract}
Spatiotemporal chaotic systems are difficult to characterize in a model-free manner because of their high dimensionality, strong nonlinearity, and sensitivity to initial conditions. Coupled map lattices, as a representative class of extended nonlinear systems, exhibit diverse regimes such as frozen random pattern, defect chaotic diffusion, and fully developed turbulence. In this work, we propose an ensemble version of multiplexing local reservoir computing for the data-driven characterization of spatiotemporal chaos. By constructing multiple base learners with randomized hyperparameters and combining their outputs, the method improves prediction robustness and quantifies predictive uncertainty through ensemble spread. More importantly, we show that this uncertainty contains direct dynamical information. It identifies frozen positions in frozen random pattern, supports the estimation of defect diffusion coefficients in defect chaotic diffusion, and provides an effective indicator of chaotic intensity in fully developed turbulence. Analyses of the spatial power spectrum and Lyapunov exponent spectrum further support the consistency between the uncertainty field and the intrinsic dynamical properties of the system. These results show that ensemble reservoir computing can serve not only as a prediction tool but also as a data-driven framework for the dynamical characterization of high-dimensional nonlinear systems.

\end{abstract}

\maketitle  

\section{Introduction}
\label{sec-intro}
Spatiotemporal chaotic systems arise in a wide range of natural and engineered settings, including climate dynamics, fluid systems, and neural activity \cite{doelman1995nonlinear,wu2024spatio,bohnstingl2022online}. Their evolution is shaped by high dimensionality, strong nonlinearity, and extreme sensitivity to initial conditions, which makes the characterization of their complex patterns and dynamical properties a fundamental challenge. As a prototypical class of extended nonlinear systems, coupled map lattices (CMLs) reproduce a variety of representative spatiotemporal behaviors, including defect chaotic diffusion, frozen random patterns, and fully developed turbulence \cite{kaneko1989spatiotemporal,kaneko1985spatiotemporal,bunimovich1988spacetime,kaneko1986collapse,kaneko1984period,crutchfield1987phenomenology}. They therefore provide an ideal testbed for the data-driven dynamical characterization of spatiotemporal chaos, including the localization of frozen structures, the estimation of defect diffusion, and the extraction of key indicators such as the spatial power spectrum and the Lyapunov exponent spectrum.

Traditional dynamical analysis methods usually rely on a priori knowledge of the governing equations or on explicit physical modeling of the system. However, for many real-world complex systems, the underlying evolution laws may be only partially known or entirely inaccessible, and first-principles modeling can become prohibitively difficult because of high dimensionality, multiscale interactions, and strong nonlinearity. Under such circumstances, data-driven approaches provide an attractive alternative for extracting dynamical information directly from observations. This perspective is particularly important for complex spatiotemporal systems, where reliable model-free characterization remains a major challenge.

Among existing data-driven approaches, reservoir computing has become a powerful tool for the analysis and prediction of nonlinear dynamical systems because of its simple training procedure, low computational cost, and strong capability for learning complex temporal dependencies \cite{jaeger2001echo,maass2002real}. By keeping the recurrent reservoir fixed and training only the output layer, reservoir computing avoids the expensive optimization required in fully trained recurrent neural networks while retaining rich nonlinear dynamics in the reservoir state. Owing to these advantages, it has been successfully applied to a wide range of tasks, including chaotic time-series prediction, system reconstruction, and the estimation of dynamical quantities from data \cite{lukovsevivcius2009reservoir,pathak2017using,lu2017reservoir,pathak2018model,lu2018attractor,pathak2018hybrid,carroll2018using,nakai2018machine,zimmermann2018observing,weng2019synchronization,griffith2019forecasting,jiang2019model,fan2020long,zhang2020predicting,guo2021transfer}. These features make reservoir computing particularly attractive for spatiotemporal nonlinear systems, where efficient learning from limited observations is often essential.

Early reservoir computing studies mainly focused on dynamical systems with relatively low-dimensional state spaces. When directly applied to high-dimensional spatiotemporal chaos, conventional RC frameworks often suffer from the curse of dimensionality and become inefficient in both training and prediction. To address this difficulty, Pathak \emph{et al.} proposed the parallel RC framework, which exploits the locality of interactions in spatiotemporal systems by decomposing the full system into multiple subdomains and training separate reservoirs for each part \cite{pathak2018model}. Following this line of development, our previous work introduced the multiplexing local reservoir computing framework \cite{lei2026symmetry}. By exploiting spatial translation symmetry, this framework aggregates local information from all lattice sites into a shared training process and enables independent prediction at each spatial position after training. For heterogeneous systems, it can be combined with parameter-aware techniques to handle quasi-translation-symmetric settings with local parameter variations \cite{kong2021machine,kong2021emergence,xiao2021predicting,panahi2024adaptable}. As a result, multiplexing local reservoir computing provides an efficient and scalable prediction framework for CML systems with homogeneous or heterogeneous local parameters, and it can be extended to other representative spatiotemporal chaotic systems such as the Kuramoto-Sivashinsky system and the Lorenz-96 model.

Although the multiplexing local reservoir computing framework achieves efficient and accurate prediction for CML systems, its deterministic output remains insufficient for the present study. In many spatiotemporal regimes, especially those with coexistence of ordered and disordered behaviors, different lattice sites may exhibit substantially different levels of predictability. A single predictor can provide only one trajectory estimate and therefore cannot quantify such heterogeneity or reveal the reliability of its own prediction across space and time. This limitation becomes particularly important when the objective extends beyond short-term forecasting to the data-driven characterization of intrinsic dynamical structures. To overcome this issue, it is necessary to equip the framework with uncertainty information that reflects variations in local predictability and can be further linked to the underlying spatiotemporal dynamics.

A natural way to introduce such uncertainty information is through ensemble learning, which improves robustness and quantifies predictive reliability by constructing multiple diverse base learners and integrating their outputs \cite{dong2020survey,sagi2018ensemble,zhou2021ensemble}. Following this idea, we incorporate an ensemble strategy into the multiplexing local reservoir computing framework and develop an uncertainty-aware approach for the data-driven characterization of spatiotemporal chaos in CML systems. Multiple base learners are generated by randomizing hyperparameters, and their outputs are combined to obtain both a robust prediction and an ensemble spread that reflects predictive uncertainty. More importantly, we show that this uncertainty is not merely a measure of prediction confidence, but also a dynamically informative quantity that reveals intrinsic features of different spatiotemporal regimes.

Specifically, in the frozen random pattern regime, the uncertainty field identifies frozen positions and localized structures. In the defect chaotic diffusion regime, it supports the estimation of defect diffusion coefficients. In the fully developed turbulence regime, it provides an effective indicator of chaotic intensity. These results show that ensemble reservoir computing can serve not only as a prediction tool but also as a data-driven framework for the dynamical characterization of high-dimensional nonlinear systems.

The remainder of this paper is organized as follows. Section 2 introduces the coupled map lattice model. Section 3 presents the ensemble reservoir computing framework. Section 4 examines the prediction performance of the ensemble reservoir computing method in the CML system. Section 5 focuses on the dynamical analysis of representative spatiotemporal regimes using the proposed framework. Finally, Section 6 summarizes the main results.

\section{Coupled Map Lattice}
\label{Two}
Coupled map lattices are prototypical models widely used in the study of spatiotemporal chaos because of their rich dynamical behaviors, including the coexistence of ordered and disordered phases \cite{kaneko1989spatiotemporal,kaneko1985spatiotemporal,bunimovich1988spacetime,kaneko1986collapse,kaneko1984period,crutchfield1987phenomenology}. A CML consists of a set of low-dimensional chaotic maps defined on lattice sites and coupled through local interactions between neighboring sites. In this work, we consider a symmetrically coupled map lattice governed by
\begin{equation}
x_{n+1}(i)=(1-\epsilon)f(x_n(i))+\frac{\epsilon}{2}\left[f(x_n(i-1))+f(x_n(i+1))\right],
\label{CMLeq}
\end{equation}
where $n$ denotes discrete time, $i$ labels the spatial site with $i=1,2,\ldots,L$, and $L$ is the total number of lattice sites. The variable $x_n(i)$ represents the local state at site $i$ and time $n$. Periodic boundary conditions are imposed such that $x_n(0)=x_n(L)$ and $x_n(L+1)=x_n(1)$. For simplicity, we adopt a coupled single-peak map lattice with local map $f(x)=1-ax^2$, where $a$ characterizes the strength of nonlinearity and $\epsilon$ denotes the coupling strength between neighboring sites.

\begin{figure*}[t]  
\centering           
\includegraphics[width=0.8\textwidth]{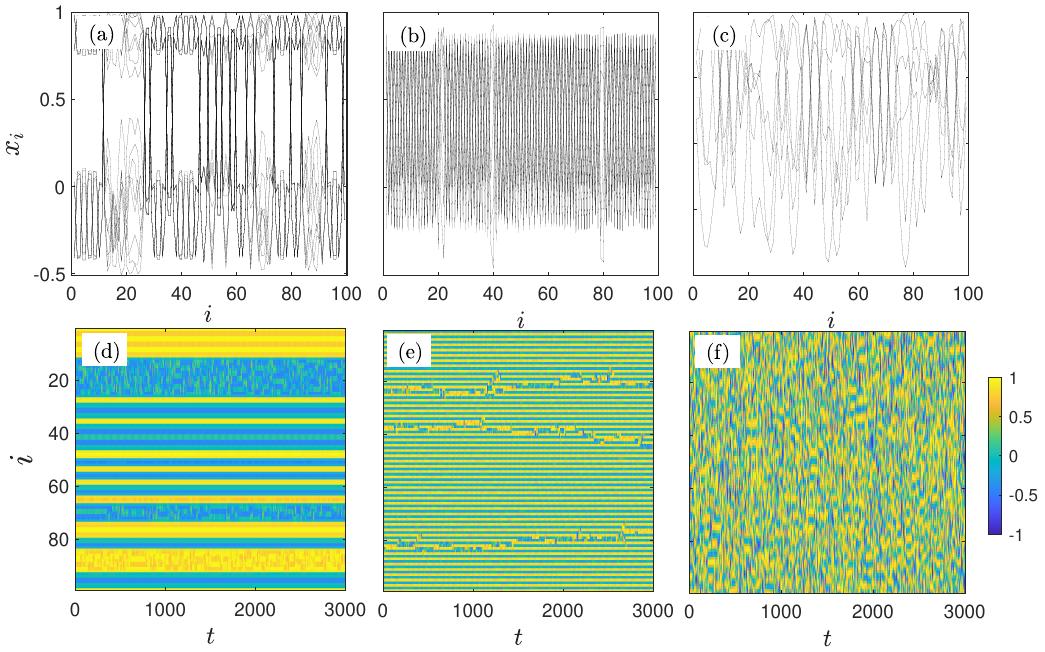} 
\setlength{\belowcaptionskip}{0cm}
\caption{The dynamics of a CML system. (a,d) (b,e) (c,f) represent spatial amplitude variation and spatio-temporal evolution with $\epsilon$=0.15,  $a$=1.52; $\epsilon$=0.1, $a$=1.79; $\epsilon$=0.2, $a$=2, respectively. } 
\label{fig1}
\end{figure*} 

The spatiotemporal evolution of the CML system is determined by the nonlinear parameter $a$ and the coupling strength $\epsilon$. Depending on their values, the system can display a variety of dynamical regimes. Among them, three representative regimes are particularly relevant to the present study, namely frozen random patterns, defect chaotic diffusion, and fully developed turbulence. Their typical spatiotemporal evolutions and corresponding snapshots are shown in Fig.~\ref{fig1}. Specifically, Figs.~\ref{fig1}(a) and \ref{fig1}(d) show a frozen random pattern, Figs.~\ref{fig1}(b) and \ref{fig1}(e) show defect chaotic diffusion, and Figs.~\ref{fig1}(c) and \ref{fig1}(f) show fully developed turbulence. In the following sections, we investigate the prediction and dynamical characterization of these three representative regimes using the proposed framework.

As the nonlinear parameter increases, the CML system undergoes a sequence of transitions from frozen random patterns to defect chaotic diffusion and then to fully developed turbulence. In the frozen random pattern regime, the system evolves into a temporally stationary but spatially disordered configuration, where some lattice sites remain effectively frozen. With further increase in the nonlinear parameter, defects emerge and move irregularly in space and time, leading to the defect chaotic diffusion regime. When the nonlinearity becomes sufficiently strong, the system enters the fully developed turbulence regime, in which irregular fluctuations spread over the entire lattice and both temporal and spatial disorder become pronounced. These three regimes provide representative examples of qualitatively distinct spatiotemporal behaviors and therefore serve as suitable benchmarks for evaluating the proposed framework.

The multiplexing local reservoir computing framework proposed in our previous work has provided an effective adaptation of reservoir computing to CML systems. By exploiting the spatial translation symmetry of the lattice, it integrates local information from all sites into a shared training process and enables independent prediction of the evolution at each spatial position. For heterogeneous CML systems with local parameter variations, it can be further combined with parameter-aware techniques to handle quasi-translation-symmetric settings and thereby achieve efficient prediction for both homogeneous and heterogeneous cases. However, in regimes with coexistence of ordered and disordered states, a single multiplexing local RC model remains insufficient. It can maintain accurate prediction over relatively long horizons for ordered nodes, but only over much shorter horizons for disordered nodes. As a result, it cannot adequately capture the heterogeneous predictability of different spatial regions, which limits its ability to support refined dynamical analysis. To overcome this limitation, we introduce an ensemble strategy into the multiplexing local RC framework and develop an uncertainty-aware extension that aims to improve robustness and enable more refined model-free dynamical characterization of typical spatiotemporal patterns in CML systems.

\section{Ensemble Reservoir Computing}
\label{Three}
Traditional single and ensemble reservoir computing frameworks both consist of three main components, namely an input module, a reservoir network, and an output module. Their operation can be divided into two stages, training and prediction. In a single reservoir computing model, prediction performance may vary because of the random initialization of the reservoir and the choice of hyperparameters. By contrast, ensemble reservoir computing employs multiple independent reservoir networks as base learners and integrates their outputs to improve robustness and overall prediction performance.

As a representative example of a low-dimensional chaotic system, we consider the Logistic map
\begin{equation}
x_{n+1}=r x_n(1-x_n),
\label{logistic}
\end{equation}
where $r$ is the control parameter and $x_n \in [0,1]$ denotes the system state at time step $n$.

When a single reservoir network is used to learn the Logistic map, the reservoir state evolves according to
\begin{equation}
\boldsymbol{r}_k =(1-\alpha)\boldsymbol{r}_{k-1}+\alpha \tanh\!\left(\boldsymbol{W}_r \cdot \boldsymbol{r}_{k-1}+\boldsymbol{W}_{in} \cdot  x_{k-1}\right),
\label{RCeq}
\end{equation}
where $\alpha$ is the leakage rate, $\boldsymbol{W}_{in}$ is the input weight matrix from the input layer to the reservoir, and $\boldsymbol{W}_r$ is the internal reservoir connectivity matrix.

The training objective is to determine an output weight matrix $\boldsymbol{W}_{out}$ such that the network output $\boldsymbol{v}_k$ approximates the target signal as accurately as possible. The output at time step $k$ is given by
\begin{equation}
\boldsymbol{v}_k=\boldsymbol{W}_{out} \cdot \boldsymbol{r}_k.
\end{equation}
The matrix $\boldsymbol{W}_{out}$ is obtained by ridge regression through minimizing the regularized loss
\begin{equation}
L=\sum_{k=T_0+1}^{T}\left\|x_k-\boldsymbol{W}_{out} \cdot \boldsymbol{r}_k\right\|^2+\beta\left\|\boldsymbol{W}_{out}\right\|^2,
\end{equation}
where $T_0$ is the length of the initial transient discarded from training, $T$ is the length of the training time series, and $\beta$ is the regularization coefficient.

For ensemble reservoir computing applied to Eq.~(\ref{logistic}), we construct $M=20$ independent base learners. The state of the $i$th base learner evolves as
\begin{equation}
\boldsymbol{r}_{k,i} =(1-\alpha)\boldsymbol{r}_{k-1,i}+\alpha \tanh\!\left(\boldsymbol{W}_{r,i} \cdot \boldsymbol{r}_{k-1,i}+\boldsymbol{W}_{in,i} \cdot x_{k-1}\right),
\label{RCeq1}
\end{equation}
where $i=1,2,\ldots,M$, and $\boldsymbol{W}_{in,i}$ and $\boldsymbol{W}_{r,i}$ denote the input and internal weight matrices of the $i$th base learner, respectively. The corresponding output weight matrices $\boldsymbol{W}_{out,i}$ are obtained independently through the same training procedure.

During the prediction stage of reservoir computing, the trained model operates in an autonomous manner. For a single RC model, the predicted output at the current step is fed back as the input for the next step, so that the future trajectory can be generated recursively. Specifically, after training, the evolution of the reservoir state in closed-loop prediction is given by
\begin{equation}
\boldsymbol{r}_{k}=(1-\alpha)\boldsymbol{r}_{k-1}+\alpha\tanh\!\left(\boldsymbol{W}_{r} \cdot \boldsymbol{r}_{k-1}+\boldsymbol{W}_{in} \cdot \hat{x}_{k-1}\right),
\end{equation}
and the corresponding prediction is
\begin{equation}
\hat{x}_{k}=\boldsymbol{W}_{out} \cdot \boldsymbol{r}_{k}.
\end{equation}

For ensemble reservoir computing, each base learner generates its own prediction sequence in the same closed-loop manner. The final ensemble prediction is then obtained by combining the outputs of all base learners. In this work, we use the ensemble mean as the final prediction and the standard deviation of the ensemble outputs as a measure of predictive uncertainty.

Let $\hat{x}_{k,i}$ denote the prediction of the $i$th base learner at time step $k$. The final ensemble prediction is defined as the average over all base learners,
\begin{equation}
\bar{x}_k=\frac{1}{M}\sum_{i=1}^{M}\hat{x}_{k,i},
\end{equation}
where $M$ is the number of base learners. To quantify the uncertainty of the ensemble prediction, we further compute the standard deviation of the ensemble outputs,
\begin{equation}
\sigma_k=\sqrt{\frac{1}{M}\sum_{i=1}^{M}\left(\hat{x}_{k,i}-\bar{x}_k\right)^2}.
\end{equation}
Here, $\bar{x}_k$ is taken as the final prediction, while $\sigma_k$ characterizes the spread among different base learners and is used as a measure of predictive uncertainty.

\subsection*{Exponential Sensitivity of Chaotic Dynamical systems}

The predictability of a chaotic system is fundamentally limited by sensitivity to initial conditions. Consider a general nonlinear system with state vector $x \in \mathbb{R}^n$ and parameters $\theta \in \mathbb{R}^p$,
\begin{equation}
\dot{x} = F(x, \theta),
\end{equation}
and let $x(t)$ be a reference trajectory starting from $x(0) = x_0$. For a small perturbation $\delta x_0$ in the initial condition, its evolution is governed by the variational equation
\begin{equation}
\frac{d}{dt} \delta x(t) = \frac{\partial F}{\partial x}\Big|_{x(t), \theta} \delta x(t),
\end{equation}
with solution
\begin{equation}
\delta x(t) = \Phi(t) \delta x_0,
\end{equation}
where $\Phi(t)$ is the fundamental solution matrix. If the system has a positive maximal Lyapunov exponent $\lambda_{\max}$, the perturbation grows approximately exponentially:
\begin{equation}
|\delta x(t)| \sim |\delta x_0| e^{\lambda_{\max} t}, \quad t \gg 1,
\end{equation}
which quantifies the exponential divergence of trajectories due to initial-condition sensitivity~\cite{galatolo2003complexity,faisst2004sensitive,geng2019second}.

Small perturbations in the system parameters, $\theta \to \theta + \delta \theta$, can induce trajectory deviations even from the same initial state $x(0)$. Denote
\begin{equation}
\delta x_\theta(t) = x(t; \theta + \delta \theta) - x(t; \theta).
\end{equation}
Linearizing for $\delta \theta \to 0$ gives the inhomogeneous variational equation
\begin{equation}
\frac{d}{dt} \delta x_\theta(t) = \frac{\partial F}{\partial x}\Big|_{x(t), \theta} \delta x_\theta(t) + \frac{\partial F}{\partial \theta}\Big|_{x(t), \theta} \delta \theta.
\end{equation}
Its solution can be expressed as
\begin{equation}
\delta x_\theta(t) = \Phi(t) \delta x_\theta(0) + \int_0^t \Phi(t-s) \frac{\partial F}{\partial \theta}\Big|_{x(s),\theta} \delta \theta \, ds.
\end{equation}
For sufficiently long times, the integral term is dominated by growth along the most unstable direction, yielding
\begin{equation}
|\delta x_\theta(t)| \sim \frac{|\delta \theta|}{\lambda_{\max}} e^{\lambda_{\max} t},
\end{equation}
where $\lambda_{\max}$ is the maximal Lyapunov exponent governing divergence due to initial-condition sensitivity. This provides a plausible theoretical basis for understanding why different base learners in ensemble RC, each corresponding to slightly perturbed parameters, diverge over time in a manner analogous to trajectories from distinct initial conditions, and why the ensemble standard deviation reflects local predictability.

For spatially extended or spatiotemporal systems, the degree of chaos can vary across both space and time. To quantify local predictability, one can compute the \emph{finite-time local Lyapunov exponent} (FT-LLE) \cite{tang1996finite,balasuriya2020uncertainty,ding2007nonlinear}. Consider a trajectory $x(t)$ at spatial location $i$ in a discrete lattice or continuous domain. The FT-LLE over a finite time interval $\Delta t$ is defined as
\begin{equation}
\lambda_i(t, \Delta t) = \frac{1}{\Delta t} \ln \frac{|\delta x_i(t + \Delta t)|}{|\delta x_i(t)|},
\end{equation}
where $\delta x_i(t)$ represents an infinitesimal perturbation at location $i$ at time $t$.

Unlike the standard (asymptotic) Lyapunov exponent, which characterizes exponential divergence in the long-time limit, the FT-LLE captures \emph{local and transient instabilities}, enabling the quantification of chaotic intensity over finite spatial and temporal scales. In many spatiotemporal systems, $\lambda_i(t, \Delta t)$ varies significantly across different spatial positions $i$, reflecting the heterogeneous nature of chaos.

Accordingly, the local predictability horizon depends on position. For a given error threshold $\Delta$, the local predictability time can be estimated as
\begin{equation}
T_i \sim \frac{1}{\lambda_i(t, \Delta t)} \ln \frac{\Delta}{|\delta x_i(t)|}.
\end{equation}
This relationship explains why some regions of a spatially extended system may remain predictable for longer times (smaller FT-LLE), while others exhibit shorter predictability (larger FT-LLE).

Such a framework provides a natural theoretical basis for understanding heterogeneous predictability in ensemble-based forecasts. In particular, it motivates the use of spatially localized measures, such as ensemble RC outputs and their associated standard deviations, to adaptively assess prediction reliability.

\subsection*{Ensemble RC for Logistic map}

\begin{figure}[htp]
\centering
\includegraphics[width=0.8\columnwidth]{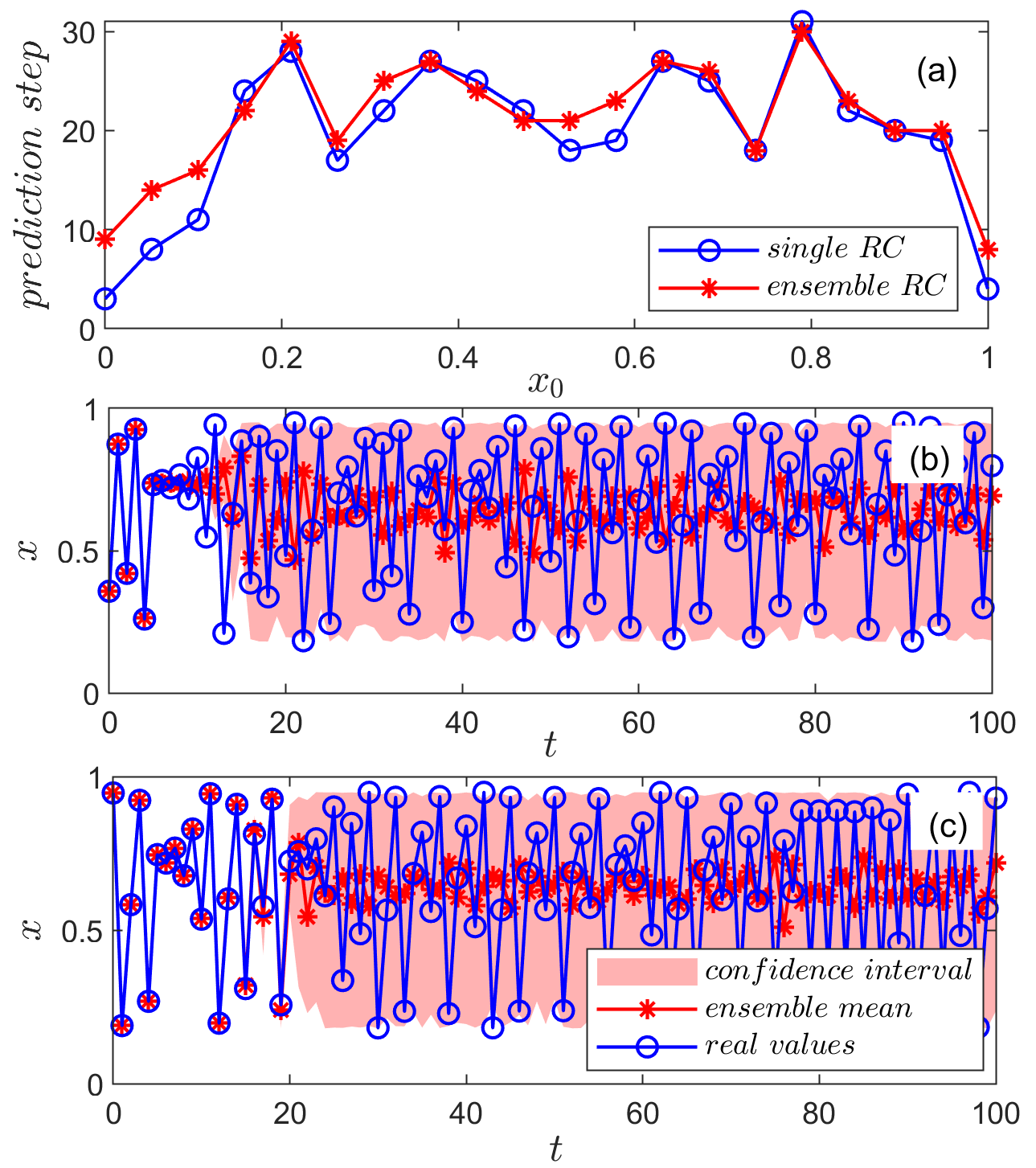}
\caption{
Prediction of the Logistic map by the ensemble reservoir computing framework. (a) Prediction horizons for trajectories starting from different initial conditions, showing strong initial-condition dependence of predictability. Both the prediction error and the ensemble standard deviation are evaluated using the same threshold value of $0.1$, which provides a unified criterion for estimating the prediction horizon. (b), (c) two representative examples with initial conditions $x_0=0.1$ and $x_0=0.5$, respectively. The blue curve denotes the true trajectory, the red curve denotes the ensemble mean, and the red shaded region denotes the uncertainty band determined by the ensemble spread.
}
\label{fig2}
\end{figure}

The ensemble standard deviation measures the spread among the predictions of different base learners and therefore provides a natural indicator of predictive uncertainty. Its practical meaning can be illustrated using the Logistic map, as shown in Fig.~\ref{fig2}.

Fig~\ref{fig2} illustrates the predictive behavior of the ensemble RC for the Logistic map from the perspective of initial-condition-dependent predictability. As shown in Fig.~\ref{fig2}(a), trajectories starting from different initial conditions exhibit markedly different prediction horizons, even though they are generated by the same dynamical system. This result indicates that the predictability of a chaotic trajectory depends strongly on the initial condition. To quantify the prediction horizon in a unified manner, we use the same threshold value of $0.1$ for both the prediction error and the ensemble standard deviation for all initial conditions. Once either the prediction error or the ensemble standard deviation exceeds this threshold, the corresponding prediction is regarded as unreliable. In this way, the ensemble standard deviation provides a convenient estimate of the effective prediction horizon and captures the large variation in predictability associated with different initial conditions.

More specific examples are shown in Figs.~\ref{fig2}(b) and \ref{fig2}(c), where two representative initial conditions, $x_0=0.1$ and $x_0=0.5$, are considered. In these panels, the blue curve denotes the true trajectory, the red curve denotes the ensemble mean, and the red shaded region represents the uncertainty band determined by the ensemble spread. For both initial conditions, the ensemble mean agrees well with the true trajectory at early prediction times, while the uncertainty band remains narrow. As the prediction proceeds, the uncertainty band gradually widens and eventually exceeds the threshold value of $0.1$, signaling the loss of predictive reliability. The two examples clearly show that different initial conditions can correspond to significantly different prediction horizons, and that the ensemble standard deviation provides an effective indicator for estimating this horizon.

\section{The prediction of the ensemble RC in the CML system} 

For spatiotemporal chaotic systems, especially those with coexistence of ordered and disordered regions, prediction difficulty can vary substantially across space and time. Although the multiplexing local reservoir computing framework has demonstrated high efficiency and good generalization ability for CML prediction tasks, deterministic prediction alone is not sufficient for the present purpose. To address this limitation, we combine the multiplexing local RC framework with ensemble learning and develop a multiplexing local RC with ensemble framework. This extension does not require any essential modification of the original reservoir architecture. Instead, it improves performance through parallel evaluation of multiple base learners. The main advantage of this ensemble strategy is that it preserves the efficient use of local information enabled by multiplexing local RC, while introducing predictive uncertainty through ensemble disagreement. As a result, the framework is both efficient and robust, and is well suited for uncertainty-aware prediction and dynamical characterization of complex spatiotemporal chaotic systems.

\begin{figure}[htp]
\centering
\includegraphics[width=0.8\columnwidth]{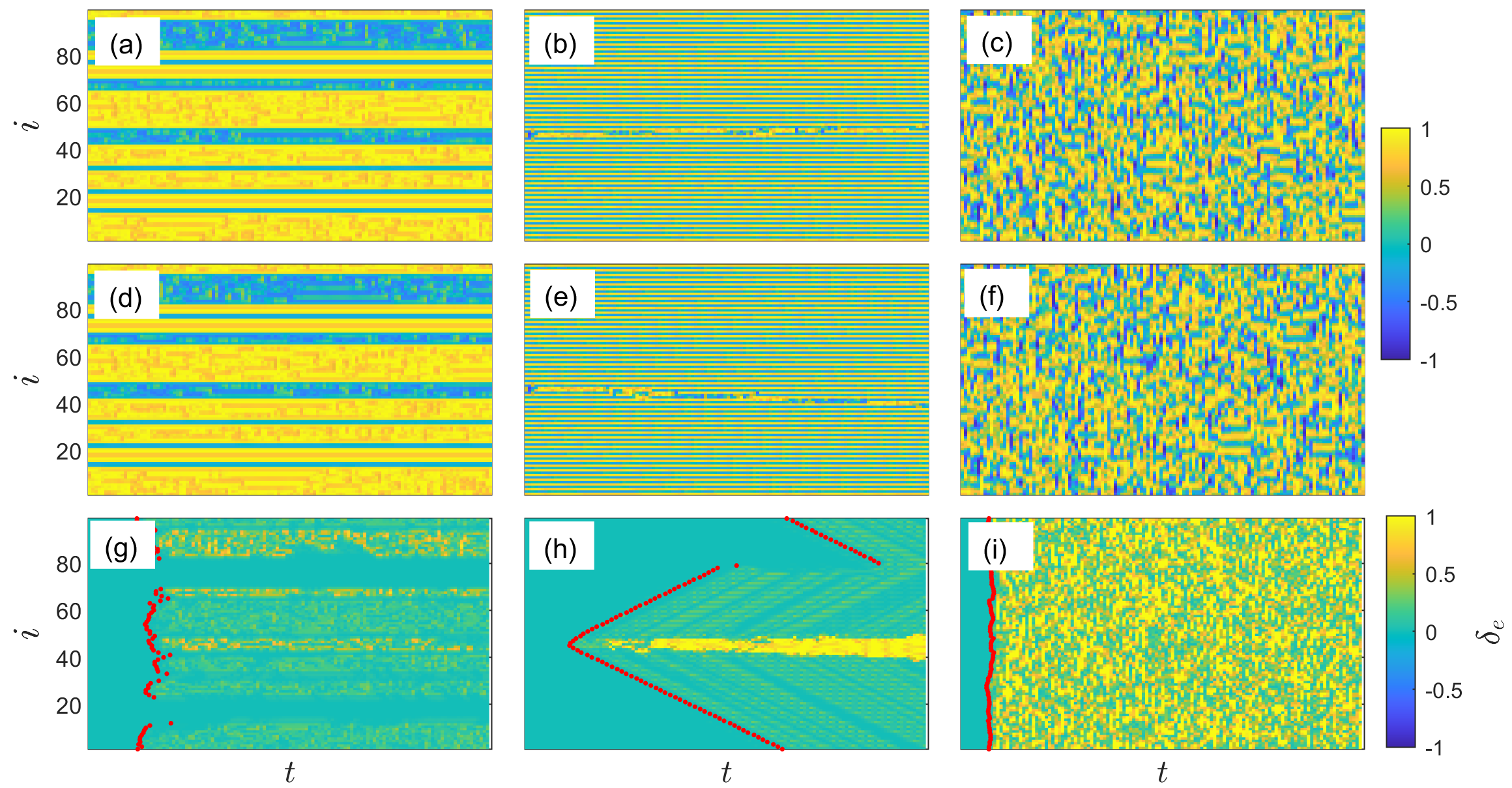}
\caption{Prediction of the CML system using the multiplexing local RC. The first three rows correspond to the true states, predicted values, error, respectively.  Panels (a, d, g), (b, e, h), and (c, f, i) represent the frozen random pattern ($\epsilon = 0.15$, $a = 1.52$), defect chaotic diffusion ($\epsilon = 0.1$, $a = 1.79$), and fully developed turbulence ($\epsilon = 0.2$, $a = 2$), respectively.}
\label{the effect of ensemble RC}
\end{figure}

Fig~\ref{the effect of ensemble RC} evaluates the performance of the multiplexing local RC framework for the symmetrically coupled CML system. The first three rows show the true states, predicted values, error, respectively. Panels (a, d, g), (b, e, h), and (c, f, i) correspond to the frozen random pattern, defect chaotic diffusion, and fully developed turbulence regimes. The standard deviation is obtained from the spread of the predictions produced by the ensemble and is used here to characterize the spatial distribution of predictive uncertainty. The red line in the third row marks the predictability limit determined by the threshold of the ensemble standard deviation. In this work, the same threshold value of $0.1$ is used uniformly, and once the ensemble standard deviation exceeds this threshold, the corresponding prediction is regarded as unreliable.

The results show that the proposed framework accurately captures the short-term evolution of the CML system in all three regimes while also reproducing their main statistical features. In the frozen random pattern regime, the method clearly resolves the spatial separation between ordered and disordered regions. In the defect chaotic diffusion regime, it correctly tracks the number and motion of defects. In the fully developed turbulence regime, it reproduces the strongly irregular spatiotemporal behavior characteristic of spatially extended chaos. These results indicate that the multiplexing local RC with ensemble is useful not only for prediction, but also for the subsequent dynamical characterization of frozen positions, defect diffusion, and turbulent chaotic intensity.

The spatial power spectrum describes how the spatial energy of the system is distributed over different wavenumbers \cite{dainty1979estimation,thebault2015statistical,jolissaint2006analytical}. It is a useful tool for analyzing the spatial modal characteristics of the CML system in the frequency domain. In this work, the spatial power spectrum is defined as
\begin{equation}
\begin{aligned}
S(k)=\left\langle \left|(1/L)\sum_{j=1}^{L} x_n(j)e^{2\pi i k j}\right|^2 \right\rangle ,
\label{Spatial power spectrum}
\end{aligned}
\end{equation}
where $j=1,2,\ldots,L$, $L$ denotes the system size, $k$ is the dimensionless wavenumber, and $\langle \cdot \rangle$ denotes the long-time average. In the numerical calculation of the spatial power spectrum, a window function is applied to the spatial sequence in order to reduce spectral leakage.

\begin{figure}[htp]
\centering
\includegraphics[width=0.8\columnwidth]{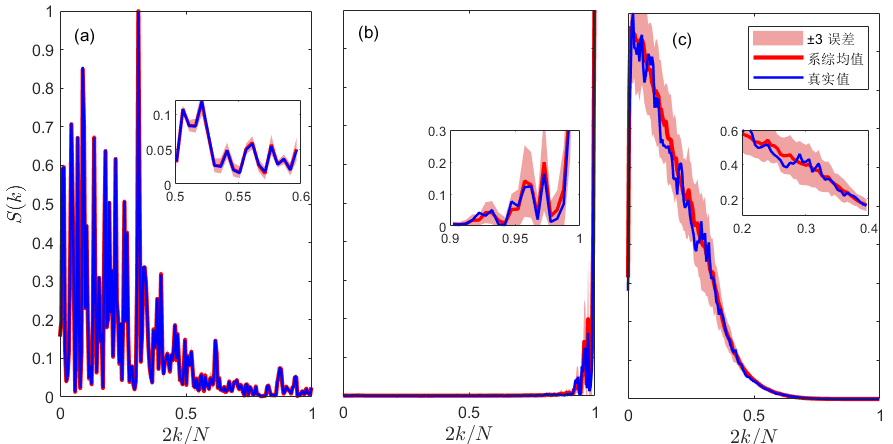}
\caption{Spatial power spectrum analysis of the CML system. The blue curves denote the spectra computed from the true states, and the red curves denote the spectra obtained from the ensemble mean prediction of the multiplexing local RC with ensemble framework. The red shaded regions indicate the uncertainty associated with the ensemble spread. Panel (a) shows the frozen random pattern with $\epsilon=0.15$ and $a=1.52$, panel (b) shows defect chaotic diffusion with $\epsilon=0.1$ and $a=1.79$, and panel (c) shows fully developed turbulence with $\epsilon=0.2$ and $a=2$.}
\label{CML_SpatialPowerSpectrum}
\end{figure}

Fig~\ref{CML_SpatialPowerSpectrum}(a-c) shows the spatial power spectra of the frozen random pattern, defect chaotic diffusion, and fully developed turbulence regimes in the CML system. The blue curves denote the spectra computed from the true CML states, while the red curves denote the spectra obtained from the ensemble mean prediction of the multiplexing local RC with ensemble framework. The red shaded regions indicate the uncertainty associated with the ensemble spread. In all three regimes, the predicted spectra agree well with the true spectra, indicating that the proposed framework captures not only the short-term evolution of the system but also its main spatial statistical features. The frozen random pattern exhibits distinct peaks, the defect chaotic diffusion regime shows coexistence of discrete peaks and a continuous background, and the fully developed turbulence regime displays a smooth continuous spectrum without pronounced discrete peaks. These results further support that the proposed framework captures the essential spatial statistical characteristics of the three regimes.

\section{Dynamical Analysis of CML with Novel RC Framework}

Beyond prediction, the multiplexing local RC with ensemble framework provides a unified tool for both forecasting and dynamical characterization of the CML system. By combining ensemble mean predictions with the associated ensemble standard deviation, key physical quantities can be extracted that are otherwise difficult to obtain directly from raw spatiotemporal data. In the following, we demonstrate its analytical capabilities across three representative regimes: frozen random patterns, defect chaotic diffusion, and fully developed turbulence.

\begin{figure}[htp]
\centering
\includegraphics[width=0.8\columnwidth]{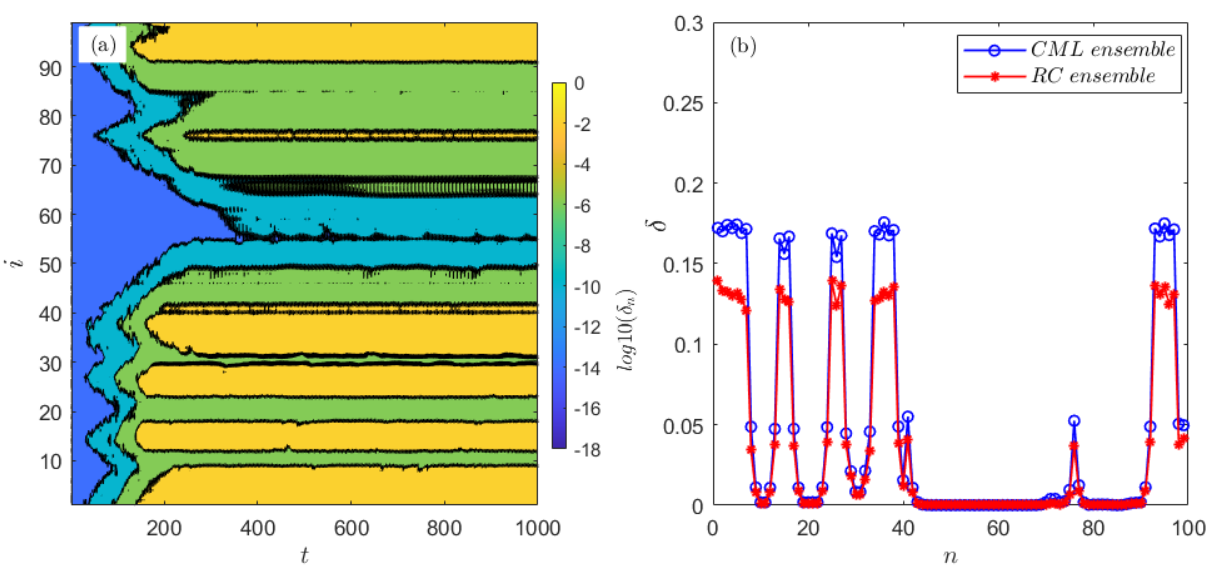}
\caption{Prediction of the frozen random pattern with $\epsilon=0.15$ and $a=1.52$. Panel (a) shows the spatiotemporal prediction by the multiplexing local RC with ensemble framework. Panel (b) compares the ensemble standard deviation $\delta$ from the true CML ensemble (blue curve) and the RC ensemble prediction (red curve) across spatial positions. The red line indicates the predictability limit using a uniform threshold of $0.1$.}
\label{frozen random pattern}
\end{figure}


In the frozen random pattern regime, the lattice is divided into spatial domains separated by kinks. Within each domain, the local dynamics may be ordered or chaotic, while the kink positions remain fixed once formed, though they depend sensitively on the initial conditions. This regime exhibits heterogeneous predictability, with ordered and disordered regions showing substantially different prediction horizons.

Fig~\ref{frozen random pattern} illustrates the ensemble RC prediction for this regime. The ensemble mean reproduces the spatiotemporal pattern accurately, while the ensemble standard deviation reflects the local variation of predictive reliability. Peaks in the standard deviation identify the boundaries of frozen structures, providing precise localization of these spatial features. The close agreement between the blue and red curves demonstrates that the framework captures the spatial stability of the system.

\begin{figure}[htp]
\centering
\includegraphics[width=0.8\columnwidth]{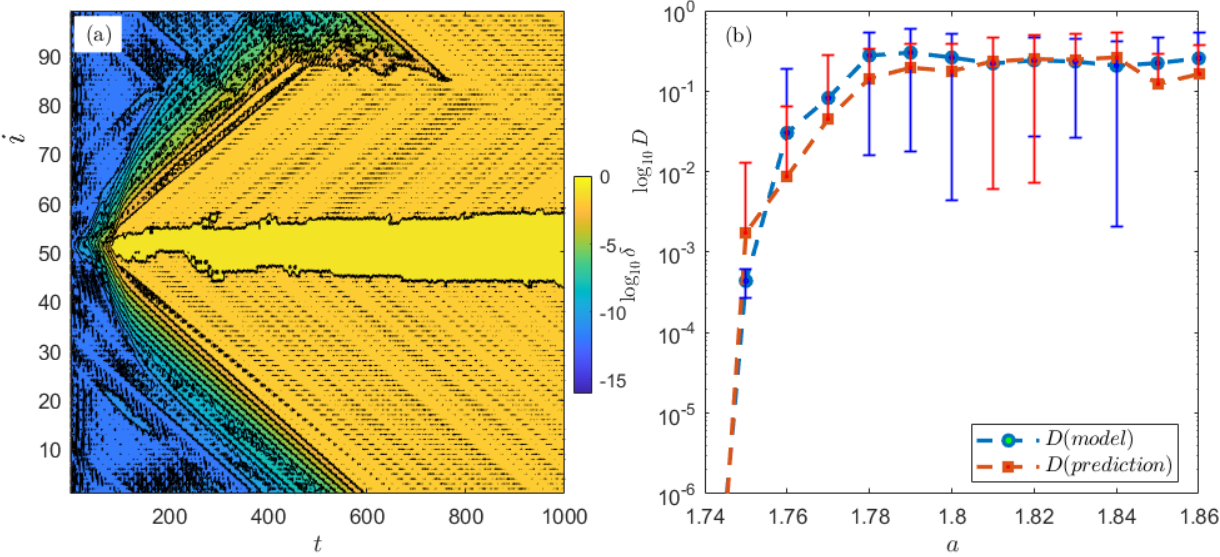}
\caption{Prediction of defect chaotic diffusion with $\epsilon=0.1$ and $a=1.79$. Panel (a) shows the spatiotemporal diagram of the ensemble standard deviation, exhibiting a parabolic diffusion pattern. Panel (b) presents the defect diffusion coefficient $D$ as a function of nonlinear parameter $a$, comparing true CML data (blue) with ensemble RC predictions (red).}
\label{diffusion_coefficient}
\end{figure}

For the defect chaotic diffusion regime, most nodes exhibit quasi-periodic behavior, while localized chaotic clusters form defects. These defects are spatially localized and move irregularly across the lattice. To quantify their dynamics, we initialize a single defect configuration:
\[
x_n(i)=x^{*}+c(-1)^n+\mathrm{rnd}(b),
\]
where $x^{*}=(\sqrt{1+4a}-1)/(2a)$, $c=0.03$, $b=0.2$, and $\mathrm{rnd}(b)$ is uniformly distributed in $[-b,b]$. The system size is $L=99$, ensuring a single defect persists. The defect position $I_n$ is identified by
\[
(x_n(i+1)-x_n(i))(x_n(i)-x_n(i-1))<0.
\]

The mean-squared displacement of the defect center is tracked for both true CML data and ensemble RC predictions, and the diffusion coefficient $D$ is obtained from
\[
\langle (I_n-I_0)^2 \rangle = 2 D n,
\]
where $\langle \cdot \rangle$ denotes an ensemble average. As shown in Fig.~\ref{diffusion_coefficient}(b), $D$ increases approximately linearly with $a$, and the ensemble RC closely follows the true data, validating the framework’s ability to quantify defect diffusion dynamics.

\begin{figure}[htp]
\centering
\includegraphics[width=0.8\columnwidth]{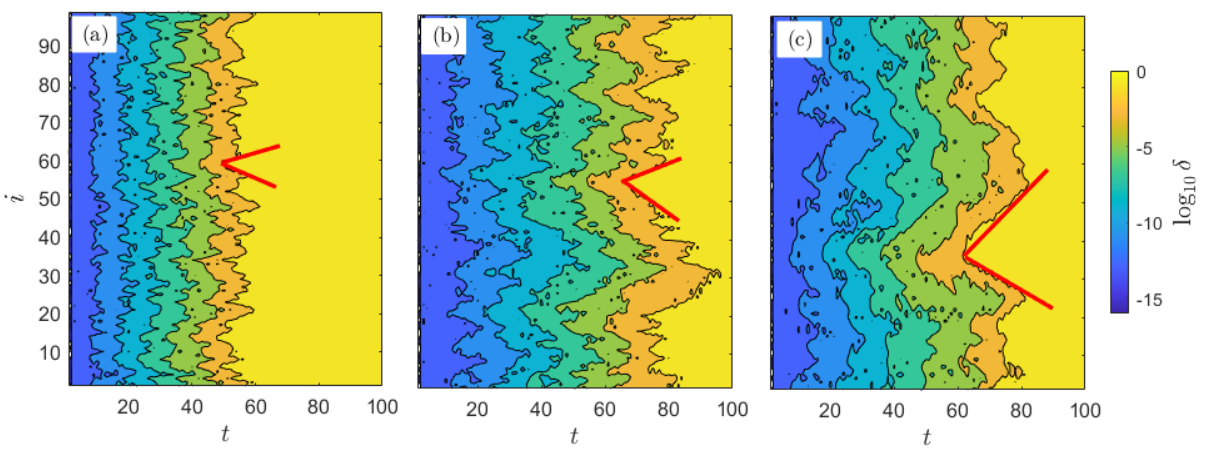}
\caption{Prediction of fully developed turbulence with $a=2$. Panels (a-c) show the spatiotemporal diagrams at $\epsilon = 0.05, 0.2, 0.75$, respectively. Ensemble standard deviation visualizes the spatial distribution of chaotic intensity.}
\label{fully developed turbulence}
\end{figure}

In the fully developed turbulence regime, the lattice exhibits strongly irregular spatiotemporal behavior with rapidly decaying spatial correlations and no clearly ordered regions. The ensemble mean accurately tracks short-term dynamics, while the ensemble standard deviation maps the growth of local unpredictability. Fig~\ref{fully developed turbulence} shows these predictions. These results demonstrate that the framework captures the global chaotic behavior and provides a quantitative measure of local unpredictability.

Across all three regimes, the ensemble multiplexing local RC framework reproduces short-term dynamics while providing spatially resolved predictive uncertainty. In the frozen random pattern, frozen structure locations are precisely identified; in defect chaotic diffusion, the defect diffusion coefficient $D$ is reliably estimated; and in fully developed turbulence, chaotic intensity across the lattice is mapped. These results confirm that the ensemble RC framework enhances predictive robustness and enables model-free extraction of meaningful dynamical information, effectively bridging the gap between accurate forecasting and spatiotemporal analysis in complex nonlinear systems.

\section{Summary}
\label{Five}
Traditional reservoir computing has demonstrated good predictive performance for general dynamical systems. However, in high-dimensional spatiotemporal chaotic systems with coexisting ordered and disordered regions, prediction horizons vary across the lattice, making it challenging to characterize the true dynamical behavior. Ensemble learning naturally addresses this challenge by leveraging the diversity of multiple base learners and complementing the properties of multiplexing local RC. By integrating ensemble strategies, the framework improves robustness and predictive accuracy while preserving efficient local information integration.

In this work, we embed ensemble learning into multiplexing local RC and propose the ensemble multiplexing local RC framework for CML prediction and analysis. The framework not only achieves stable short-term predictions of the spatiotemporal sequence but also provides a spatially resolved measure of predictive uncertainty through the ensemble mean and standard deviation. Across the three representative CML regimes—frozen random patterns, defect chaotic diffusion, and fully developed turbulence—the framework reproduces key dynamical features. It precisely identifies frozen structures, accurately estimates defect diffusion coefficients, and quantifies chaotic intensity, confirming its effectiveness in capturing both local and global nonlinear behaviors.

These results demonstrate that ensemble multiplexing local RC extends traditional reservoir computing beyond prediction, enabling model-free extraction of meaningful dynamical information from uncertainty distributions. The approach offers a new perspective and a practical tool for analyzing spatiotemporal chaos. Looking forward, the framework can be extended to other classical spatiotemporal chaotic systems, such as coupled Lorenz lattices, Kuramoto-Sivashinsky equations, and reaction-diffusion systems, providing a generalizable method for exploring the dynamics of complex nonlinear systems.

\section*{Acknowledgment}
This work was supported by the National Natural Science Foundation of China (NNSFC) under Grant No.~62333002 and Opening Project of State Key Lab of Information Photonics and Optical Communications under Grant No. IPOC2023ZJ02, and the Fundamental Research Funds for the Central Universities under Grant No. 2024RC11.

\bibliography{references1}

\end{document}